\documentclass[twocolumn,           
               showpacs,            
               preprintnumbers,     
               aps,                 
               prd,          	    
               a4paper,             
               superscriptaddress,  
               nofootinbib,         
               tightenlines,        
               floats,floatfix      
               ]{revtex4}

\usepackage{graphicx}
\usepackage{latexsym}
\usepackage{amsmath,amssymb}        
\usepackage[draft=false]{hyperref}

\begin{document}



\title{How long before the end of inflation were observable perturbations 
produced?}
\author{Andrew R.~Liddle}
\affiliation{Astronomy Centre, University of Sussex, Brighton BN1 9QJ, United 
Kingdom}
\author{Samuel M.~Leach}
\affiliation{D\'epartement de Physique Th\'eorique, Universit\'e de Gen\`eve,
24 quai Ernest Ansermet, CH-1211 Gen\`eve 4, Switzerland}
\date{\today}
\pacs{98.80.Cq \hfill astro-ph/0305263}
\preprint{astro-ph/0305263}


\begin{abstract}
We reconsider the issue of the number of $e$-foldings before the end of 
inflation at which observable perturbations were generated. We determine a 
plausible upper limit on that number for the standard cosmology which is around 
60, with the expectation that the actual value will be up to 10 below this. We 
also note a special property of the $\lambda \phi^4$ model which reduces the 
uncertainties in that case and favours a higher value, giving a fairly definite 
prediction of 64 $e$-foldings for that model. We note an extreme (and highly 
implausible) situation where the number of $e$-foldings can be even higher, 
possibly up to 100, and discuss the shortcomings of quantifying inflation by 
$e$-foldings rather than by the change in $aH$. Finally,
we discuss the impact of 
non-standard evolution between the end of inflation and the present, showing 
that again the expected number of $e$-foldings can be modified, and in some 
cases significantly increased.
\end{abstract}

\maketitle


\section{Introduction}

With observations of perturbations in the Universe reaching a quality that 
seriously constrains inflationary models \cite{Petal}, it is timely to revisit 
one of the 
significant uncertainties in fixing the inflationary model, being the location 
on the inflationary potential corresponding to the observed perturbations. This 
is usually quantified by the number of $e$-foldings before the end of inflation 
at which our present Hubble scale equalled the Hubble scale during inflation --- 
the epoch of horizon crossing. While in most inflation models the spectrum of 
perturbations generated depends only on the dynamics of the Universe around 
horizon crossing, determination of the number of $e$-foldings requires a model 
of the entire history of the Universe.

Determining the appropriate number of $e$-foldings may shed light on the 
mechanism ending inflation (a goal that would also be greatly assisted by a 
determination of the energy scale of inflation). There are currently two popular 
mechanisms: 
steepening of the potential leading to an end of the slow-roll era, or the 
hybrid inflation mechanism where an instability in a second field brings 
inflation to an end. In the latter case, the number of $e$-foldings does not 
have great significance, but in the case of slow-roll violation, it is a 
significant constraint on the inflationary potential that inflation must come to 
an end a particular number of $e$-foldings after the observed perturbations were 
generated. It is desirable to combine this constraint with those coming from the 
form of the observed perturbations.

In this paper we revisit the issue of the number of $e$-foldings, highlighting
the sources of uncertainty.  In particular, we seek to impose robust upper and
lower limits on the number of $e$-foldings corresponding to observable
perturbations, both in the case of the standard cosmological history and for 
models with different early evolution of the Universe.\footnote{Our results say 
nothing about the total number of $e$-foldings which may have taken place, which 
is expected to be much larger.}  We are able to make some technical improvements 
to previous
calculations, now that the Standard Cosmological Model, featuring a low-density
spatially-flat Universe, is firmly established.  Further, we are able to
investigate how the number of $e$-foldings is modified as one changes the
properties of inflation models within the range allowed by observations.

As we were completing this paper, a paper appeared by Dodelson and Hui 
\cite{DH}, 
who also consider the maximum number of $e$-foldings of inflation but with a 
less wide-ranging treatment than ours. While the original version of their paper 
had some discrepancies as compared to ours, they submitted a revised version of 
their paper simultaneously with ours which is in good agreement where the 
discussion overlaps.

\section{The simplest cosmology}

Our main aim is to obtain the number of $e$-foldings $N(k)$ before the end of 
inflation at which a comoving scale $k$ equalled the Hubble scale $aH$. Normally 
we will focus on the scale $k_{{\rm hor}} = a_0 H_0$ which equals the present 
Hubble 
scale. Current observations are able to probe from around this scale up to $k$ 
values about three orders of magnitude larger using microwave anisotropy and 
galaxy clustering data, and perhaps a further order of magnitude using quasar 
absorption line features, corresponding to a range of about 10 $e$-foldings in 
total.

The number of $e$-foldings during inflation, $N(k)$, is defined by
\begin{equation}
e^{N(k)} \equiv \frac{a_{{\rm end}}}{a_k} \,,
\end{equation}
where $a_{{\rm end}}$ is the value of the scale factor at the end of inflation 
and $a_{{\rm k}}$ is its value when the scale $k$ equalled $aH$ during 
inflation.\footnote{As discussed by Liddle, Parsons and Barrow \cite{LPB}, it 
makes more logical sense to define the amount of inflation as the ratio 
of $aH$, rather than $a$. More on that later; for now we follow the standard 
usage.} We will use $N_{{\rm hor}}$ to indicate $N(a_0 H_0)$.

\begin{figure}[t]
\includegraphics[scale=0.38,angle=0]{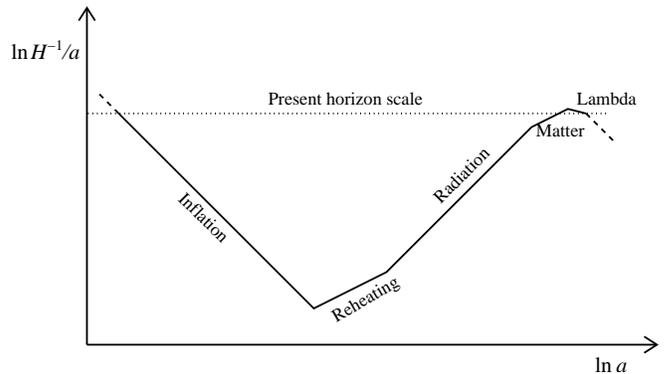}
\caption{A plot of $\ln(H^{-1}/a)$ versus $\ln a$ shows the different epochs in 
the $e$-foldings calculation. The solid curve shows the evolution from the 
initial horizon crossing to the present, with the dashed lines showing likely 
extrapolations into the past and future.
The condition for inflation is that $\ln(H^{-1}/a)$ be decreasing. Lines of 
constant Hubble parameter (not shown) lie at 45 degrees (running top left to 
bottom right). The 
limit of exponential inflation gives a line at this angle, otherwise the 
inflation line is shallower. 
During reheating and matter domination $H^{-1}/a \propto a^{1/2}$, while during 
radiation domination $H^{-1}/a \propto a$. The recent domination by dark energy 
has initiated a new era of inflation. The horizontal dotted line indicates the 
present horizon scale. The number of $e$-foldings of inflation is the horizontal 
distance 
between the time when $H^{-1}/a$ first crosses that value and the end of 
inflation.}
\label{f:schem}
\end{figure}

To determine the number of $e$-foldings corresponding to a scale measured in 
terms of the present Hubble scale, we need a complete model for the history of 
the Universe. At least from nucleosynthesis onwards, this is now well in place, 
but at earlier epochs there are considerable uncertainties. At this stage, we 
make the following simple assumptions for the sequence of events after 
inflation, considering possible alternatives in the next section. We assume 
that inflation is followed by a period of reheating, during which the Universe 
expands as matter dominated (this assumption is not true in all models --- see 
subsection~\ref{s:quartic}). This then gives way to a period of radiation 
domination, which according to the Standard Cosmological Model lasts until a 
redshift of a few thousand before giving way to matter domination, and then 
finally at a redshift below one to a cosmological constant or quintessence 
dominated era.
We assume sudden transitions between these epochs, labelling the end of the 
reheating period by `reh' and the matter--radiation equality epoch by `eq'. This 
is illustrated in Figure~\ref{f:schem}.

We can therefore write
\begin{equation}
\frac{k}{a_0 H_0} = \frac{a_k H_k}{a_0 H_0} = e^{-N(k)} \, \frac{a_{{\rm 
end}}}{a_{{\rm reh}}} \, \frac{a_{{\rm reh}}}{a_{{\rm eq}}} \, 
\frac{H_k}{H_{{\rm eq}}} \,\frac{a_{{\rm eq}}H_{{\rm eq}}}{a_0 H_0} \,
\end{equation}
Some useful factors are (see e.g.~Ref.~\cite{LL})
\begin{eqnarray}
\frac{a_{{\rm eq}}H_{{\rm eq}}}{a_0 H_0} & = & 219 \, \Omega_0 h \,; \\
H_{{\rm eq}} & = & 5.25 \times 10^6 \, h^3 \, \Omega_0^2 H_0 \,;\\
H_0 & = & 1.75 \times 10^{-61} \, h \, m_{{\rm Pl}} \quad \mbox{with $h \simeq 
0.7$}
\end{eqnarray}
Using the slow-roll approximation during inflation to write $H_k^2 \simeq 8\pi 
V_k/3m_{{\rm Pl}}^2$, we obtain
\begin{eqnarray}
\label{e:efolds}
N(k) & = &  -\ln \frac{k}{a_0 H_0} + \frac{1}{3} \ln \frac{\rho_{{\rm 
reh}}}{\rho_{{\rm end}}} + \frac{1}{4} \ln \frac{\rho_{{\rm eq}}}{\rho_{{\rm 
reh}}}  \nonumber \\
&& \quad \quad \quad+ \ln \sqrt{\frac{8 \pi V_k}{3 m_{{\rm Pl}}^2}} \, 
\frac{1}{H_{{\rm eq}}} + \ln 219 \Omega_0 h \,.
\end{eqnarray}
which agrees with Refs.~\cite{LLrep,LL} while being more precise about the 
prefactor. In fact ultimately the dependence on the matter density $\Omega_0$ 
will cancel 
out, and though a dependence on $h$ remains this parameter is now accurately 
determined by observations.

\subsection{A plausible upper limit}

\label{s:plaus}

The evolution of the Universe as described above is a plausible model for its 
entire history. Nevertheless, there are significant uncertainties in applying 
Eq.~(\ref{e:efolds}). $V_k$ is a quantity we would hope to extract from the 
perturbations, but presently only upper limits exist, as the density 
perturbation amplitude depends on a combination of the potential and its slope, 
being unable to constrain either separately. Detection of primordial 
gravitational waves, which so far has not been achieved, is needed to break this 
degeneracy. We do not know how prolonged the reheating epoch might be, which is 
needed to determine $\rho_{{\rm reh}}$, nor how much lower the energy density 
$\rho_{{\rm end}}$ at the end of inflation might be as compared to $V_k$.

Nevertheless, we can impose a plausible maximum on the number of $e$-foldings by 
making an assumption, namely that there is no significant drop in energy density 
during these last stages of inflation, so that $V_k = \rho_{{\rm end}}$. Note 
however that this is {\em not} the correct way to maximize Eq.~(\ref{e:efolds}), 
a topic we return to in subsection~\ref{s:maximize}, and so is a non-trivial 
assumption. Having made it, the inflation line in Figure~\ref{f:schem} lies at 
45 degrees, and we can maximize the number of $e$-foldings by 
assuming 
that reheating is instantaneous, so 
that $\rho_{{\rm reh}}=\rho_{{\rm end}}$. Focussing now on the current horizon 
scale, this gives a maximum number of $e$-foldings corresponding to the horizon 
scale of 
\begin{equation}
N_{{\rm hor}}^{{\rm max}} =  \frac{1}{4} \ln \frac{\rho_{{\rm eq}}}{V_{{\rm 
hor}}} + \ln \sqrt{\frac{8 \pi V_{{\rm hor}}}{3 m_{{\rm Pl}}^2}} \, 
\frac{1}{H_{{\rm eq}}} + \ln 219 \Omega_0 h \,,
\end{equation}
and substituting in the known quantities gives
\begin{equation}
N_{{\rm hor}}^{{\rm max}} = 68.5 + \frac{1}{4} \ln \frac{V_{{\rm 
hor}}}{m_{{\rm Pl}}^4} \,.
\end{equation}

The potential energy is bounded by the requirement that perturbations have the 
observed amplitude. For the accuracy level currently required, we can assume 
that the perturbations are entirely from density perturbations, whose amplitude 
is given in the slow-roll approximation by \cite{LL}
\begin{equation}
{\cal P}_{{\rm S},0} = \frac{8V}{3m_{{\rm Pl}}^4} \frac{1}{\epsilon} \,,
\end{equation}
where
\begin{equation}
\epsilon = \frac{m_{{\rm Pl}}^2}{16\pi} \, \left( \frac{dV/d\phi}{V} \right)^2 
\,,
\end{equation}
is the usual slow-roll parameter which observations restrict to $\epsilon 
\lesssim 0.05$. The observed perturbation amplitude on large scales is ${\cal 
P}_{{\rm S},0} \simeq 2.6 \times 10^{-9}$ \cite{LeaL} (ignoring a weak 
dependence on the precise form of the perturbations generated), giving
\begin{equation}
\label{e:max}
N_{{\rm hor}}^{{\rm max}} = 63.3 + \frac{1}{4} \ln \epsilon \,.
\end{equation}
A similar formula was obtained by Dodelson and Hui \cite{DH} who additionally 
imposed an upper limit on $\epsilon$ from gravitational wave limits.
Note that in some models of inflation, particularly hybrid inflation models, 
$\epsilon$ can be very small indeed; enough to make the last factor significant.

We have analyzed the values of $N_{{\rm hor}}^{{\rm max}}$ for elements of a 
Monte Carlo Markov Chain fit to a set of observational data including WMAP and 
2dFGRS, which generates values of $V$ and the slow-roll parameters directly from 
the data as described in Ref.~\cite{LLMC}. 
This confirms that for single-field inflation models the dependence on higher 
slow-roll parameters (via 
the changed normalization) is negligible and that Eq.~(\ref{e:max}) is an 
excellent description.

The formula we used for the perturbation amplitude assumes that there is only 
one dynamically-important field during inflation, and may be modified if 
multi-field effects are important -- see Ref.~\cite{multi}. It would require a 
very large change in the perturbation amplitude to make a significant difference 
to Eq.~(\ref{e:max}), but if such a dramatic change is expected in a particular 
model, it would be necessary to recalculate the number of $e$-foldings 
specifically for that case.

We conclude that a plausible maximum number of 
$e$-foldings that can correspond to observable scales is around 62 for the 
standard picture of cosmological evolution following inflation. We stress that 
this says nothing about the total number of $e$-foldings that take place, which 
is expected to be much larger.

\subsection{A standard hypothesis}

The assumptions made in the last subsection are not expected to hold precisely, 
and 
hence the expected number of $e$-foldings will be different. In this subsection, 
we 
assess how different the number is expected to be, while remaining in the 
framework of the simplest cosmological history.

The two effects we need to allow for are that $\rho_{{\rm end}}$ will be less 
than $V_{{\rm hor}}$, and that $\rho_{{\rm reh}}$ will be less than $\rho_{{\rm 
end}}$. We can write
\begin{equation}
\label{e:standard}
N_{{\rm hor}} = N^{{\rm max}}_{{\rm hor}} + \frac{1}{4} \ln \frac{V_{{\rm 
hor}}}{\rho_{{\rm end}}} + \frac{1}{12} \ln \frac{\rho_{{\rm reh}}}{\rho_{{\rm 
end}}} \,.
\end{equation}
The former effect is the one neglected in the previous subsection. Note that it 
{\em increases} the number of $e$-foldings required, an effect we study fully in 
subsection~\ref{s:maximize}. 
In hybrid inflation 
models, it is expected that there is very little reduction in the energy density 
during the late stages, while in slow-roll inflation models the reduction is 
typically one or two orders of magnitude. This term is therefore unlikely to 
increase $N_{{\rm hor}}$ by much more than one.

The main uncertainty resides in the final term. Reheating can easily be a 
prolonged process, during which the energy density drops by orders of magnitude. 
Indeed, in supersymmetric theories avoidance of overproduction of gravitinos 
requires an energy density below $(10^{11} \, {\rm GeV})^4$ \cite{KM}, implying 
a drop in energy density of around twenty orders of magnitude unless $\epsilon$ 
has a tiny value. The most extreme assumption would be that reheating continues 
almost to nucleosynthesis, giving a lower limit at about $(10^{-3} \,  {\rm 
GeV})^4$, though usually the electroweak scale $(10^{2} \,  {\rm GeV})^4$ is 
regarded as the practical limit. Luckily the dependence has a prefactor of 
$1/12$, so those three energy scales correspond to a reduction of $N_{{\rm 
hor}}$ by only 4, 15 and 11 respectively for the case of large $\epsilon$. These 
numbers can be reduced if $\epsilon$ is tiny as then the inflationary energy 
scale will be lower, but then a similar correction will be accrued from the $\ln 
\epsilon$ term in Eq.~(\ref{e:max}). However the gravitino limit may not apply 
in all models. In summary, a plausible value for the reduction in $N_{{\rm 
hor}}$ caused by reheating is 5 $e$-foldings, with a likely range of about 5 in 
either direction around that. 

Putting that information together, in the context of the simplest cosmology, a 
reasonable fiducial value for the number of $e$-foldings corresponding to the 
present Hubble scale is around 55, with an uncertainty of 5 around that. In the 
literature values of 50 or 60 are common, and in fact lie towards the extremes. 
However we will see that, under fairly reasonable assumptions, there are several 
ways in which the number of $e$-foldings could lie outside that range, in either 
direction.

\subsection{The special case of $\lambda \phi^4$}

\label{s:quartic}

The quartic potential $V = \lambda \phi^4$ has been of particular interest 
lately as it lies in the region excluded by the WMAP analysis \cite{Petal}. As 
the precise predictions for the spectra depend on the number of $e$-foldings, 
some care is required with models which are close to the exclusion limit, as 
highlighted by Barger et al.~\cite{BLM}. 

It turns out that for $\lambda \phi^4$ we can be more precise, because 
reheating in a quartic potential has an unusual property --- the expansion 
during the scalar field oscillations is as radiation dominated \cite{ford}, 
rather than the matter-dominated expansion given by oscillations in a quadratic 
potential. Accordingly, the duration of the epoch of reheating no longer matters 
and we can take the Universe as radiation-dominated beginning at the end of 
inflation.\footnote{This picture may be altered if there is significant 
preheating \cite{preh}. However usually it is assumed that the particles 
produced by preheating are rapidly converted to radiation, in which case the 
result as described in unchanged. If more complicated preheating phenomenology 
takes place (e.g.~as in Ref.~\cite{infpreh}) our results may be modified.} This 
gives
\begin{equation}
\label{e:quartic}
N_{{\rm hor}}^{{\rm quartic}} = N^{{\rm max}}_{{\rm hor}} + \frac{1}{4} \ln 
\frac{V_{{\rm hor}}}{\rho_{{\rm end}}} \,.
\end{equation}

Additionally, as we have a definite model we can compute the ratio $V_{{\rm 
hor}}/\rho_{{\rm end}}$, which the slow-roll approximation gives as (see 
e.g.~Ref.~\cite{LL})
\begin{equation}
\frac{V_{{\rm hor}}}{\rho_{{\rm end}}} \simeq N^2 \,,
\end{equation}
and the value of $\epsilon$ which is $1/N$. Putting all this together gives
\begin{equation}
N_{{\rm hor}}^{{\rm quartic}} = 63.3 + \frac{1}{4} \ln N_{{\rm hor}} \,,
\end{equation}
whose solution is $N_{{\rm hor}}^{{\rm quartic}} = 64$.
Hence under the assumptions of the simplest cosmology, the quartic potential 
allows an accurate specification of the number of $e$-foldings, the only 
approximation being the assumption of instantaneous transitions between epochs. 
The value in this model is unusually high, because of the non-standard behaviour 
during reheating and the significant reduction in $H$ during the late stages 
which leads to it violating the limit of the previous subsection. This large 
value means that the model is around the borderline of what present data allows 
\cite{Petal,BLM,DH,LLMC}

\subsection{Extreme cases, and a better definition of inflation}

\label{s:maximize}

As the $\lambda \phi^4$ case has illustrated, the plausible upper limit of 
subsection~\ref{s:plaus} is not as rigid as one would like, because reduction of 
the energy scale during inflation can play an important role. What inflation is 
really trying to achieve is to increase the ratio $aH$, and every reduction in 
$H$ by a factor $e$ then requires an extra $e$-folding of expansion to counter 
it. In terms of Figure~\ref{f:schem}, the inflation line is shallower, and 
hence has a greater horizontal extent before reaching the standard 
post-inflationary evolution. Although the 
reduction in energy density shortens the 
evolution after inflation, it is clear from the figure (or inspection of 
Eq.~(\ref{e:efolds})) that reducing the energy density during inflation wins, 
with the largest possible $N$ 
being given by the smallest possible $\rho_{{\rm end}}$ (if all other parameters 
are unchanged) accompanied by instantaneous reheating. This again gives us 
Eq.~(\ref{e:quartic}), and in absolute generality $\rho_{{\rm end}}$ could be as 
late as nucleosynthesis, giving
\begin{equation}
N_{{\rm hor}}^{{\rm extreme}} \simeq 107 \,.
\end{equation}

This is a surprisingly large value, and no plausible inflation model will 
generate it, but we mention it as possible in principle. To achieve such a large 
reduction in energy density while inflating, inflation must take place extremely 
close to the `coasting' limit of $a \propto t$, at which $aH$ remains constant. 
In that limit, the $e$-foldings of inflation are very inefficient at pushing 
scales $k$ outside the horizon $aH$. Note that such an evolution is not possible 
on scales with observable perturbations, as the generated spectrum would be far 
from scale-invariant, but nothing in principle stops it occurring at the later 
stages. 

A concrete example would be as follows. At a high energy scale, say $V_{{\rm 
hor}} = (10^{16} \, {\rm GeV})^4$, we have a typical inflationary expansion, 
generating nearly scale-invariant perturbations and pushing them around 20 
$e$-foldings outside the horizon. This epoch then gives way to a fast-rolling 
inflationary epoch\footnote{See Ref.~\cite{fast} for a general discussion of 
fast-roll inflation.} with $a \propto t^p$ where $p$ only slightly exceeds 1, 
with 
this fast-rolling epoch continuing all the way down to $\rho_{{\rm end}} = (1 \, 
{\rm GeV})^4$. During the fast-roll era the perturbation spectrum will 
have sharply decreasing amplitude. As the density during this fast-roll stage is 
$\rho \propto 1/a^{2/p}$, this generates a further $\Delta N = (2/p) \ln 
(V_{{\rm hor}}/\rho_{{\rm end}}) = 72/p \simeq 72$ $e$-foldings. As during this 
evolution $aH \propto t^{p-1} \simeq {\rm constant}$, scales are not pushed 
further outside the Hubble radius during the fast-roll epoch, and so the 
perturbations generated during the slow-roll phase are correctly positioned to 
be those observable at the present epoch, even though nearly 100 $e$-foldings 
have taken place since they were generated. 

The issues raised in this subsection would be completely 
avoided had the more logical definition of the amount of inflation as the change 
in $\tilde{N} \equiv \ln(aH)$ been used \cite{LPB}. This definition 
automatically 
accounts for the reduction in $H$ during inflation, and is given by
\begin{equation}
\tilde{N}(k) = N(k) + \ln \frac{H_{{\rm end}}}{H_k} = N(k) - \frac{1}{2} \ln 
\frac{V_k}{\rho_{{\rm end}}} \,.
\end{equation}
This is sufficient to change the sign of the troublesome coefficient in 
Eqs.~(\ref{e:standard}) and (\ref{e:quartic}), thus ensuring that $\tilde{N}$ is 
maximized by taking the largest possible $\rho_{{\rm end}}$ and $\rho_{{\rm 
reh}}$. The plausible upper limit of Eq.~(\ref{e:max}) would then apply in 
general to $\tilde{N}$, including in the case of the quartic potential where 
$\tilde{N}$ is significantly less than $N$.

\section{Non-standard cosmologies: upper and lower limits}

\label{s:complex}

The previous section considered only the case of the simplest cosmology, where 
inflation gives way to reheating and then to the standard Hot Big Bang 
evolution. However the appropriate value for $N_{{\rm hor}}$ is sensitive to 
modifications to that assumption, and there are no direct constraints on the 
evolution for most of the early history of the Universe.

In general these modifications could either increase or decrease $N_{{\rm 
hor}}$. The two modifications we discuss which are restricted to the period 
after inflation both serve to reduce the value of $N_{{\rm hor}}$. However we 
also discuss two possibilities which can raise $N_{{\rm hor}}$, though both 
require 
modifications to the way inflation is modelled.

In this section, we will neglect the possibility of a significant reduction in 
the energy density during the last stages of inflation, though such a reduction 
should be combined with the effects discussed here whenever a definite model is 
under discussion, and could be conveniently addressed by use of $\tilde{N}$ in 
place of $N$.

\subsection{An upper limit}

\label{s:except}

Although Section~\ref{s:plaus} gives a plausible upper limit to the number of
$e$-foldings for inflation assuming roughly constant energy density, it is still 
possible to raise the number further.  What is needed is to
replace part of the radiation-dominated era with a period where the Universe
expands even more slowly.  The limiting case consistent with causality is a
stiff fluid dominated era where $p = \rho$, giving $a \propto t^{1/3}$ and $\rho
\propto 1/a^6$.  In fact, such a period is not at all ridiculous, as this is the
expansion law for a kinetic-energy dominated scalar field, and the literature
contains several proposals for ending inflation by the inflaton field making a
transition from potential energy domination to kinetic energy domination
\cite{kin}.  Further, such kinetic energy dominated periods tend to be prolonged
if reheating is to proceed by gravitational particle production
\cite{ford,gravpart}.

Instead of conventional reheating, we will consider a stiff fluid to dominate 
until an energy density $\rho_{{\rm kin}}$, before giving way to radiation 
domination as before. Considering 
Eq.~(\ref{e:efolds}), the effect is to make the replacement
\begin{equation}
\frac{1}{3} \ln \frac{\rho_{{\rm reh}}}{\rho_{{\rm end}}} +
\frac{1}{4} \ln \frac{\rho_{{\rm eq}}}{\rho_{{\rm reh}}} \longrightarrow 
\frac{1}{6} \ln \frac{\rho_{{\rm kin}}}{\rho_{{\rm end}}} + \frac{1}{4} \ln 
\frac{\rho_{{\rm eq}}}{\rho_{{\rm kin}}} \,.
\end{equation}
In order to find out how large this effect could be on the maximum number of 
$e$-foldings, we again take $\rho_{{\rm reh}}=\rho_{{\rm end}}$ for the original 
scenario, while in the new scenario we take $\rho_{{\rm kin}}$ to be as small as 
possible. The Universe must have attained thermalized radiation domination by 
the time of 
nucleosynthesis, so the most radical modification is for the kinetic regime to 
end shortly before nucleosynthesis, at $\rho_{{\rm nuc}} \simeq (10^{-3} \, {\rm 
GeV})^4$. The possible increase in $N$ is therefore
\begin{equation}
N_{{\rm extra}} = \frac{1}{12} \ln \frac{\rho_{{\rm end}}}{\rho_{{\rm nuc}}}
\end{equation}
As $\rho_{{\rm end}}$ could 
be as high as $(10^{16} \, {\rm GeV})^4$, in the most extreme case 
this can increase the number of $e$-foldings by as much as 15, as compared to 
the plausible maximum of Section~\ref{s:plaus}.

In fact, stuff fluid cosmologies are constrained by the possibility of an 
excessive gravitational wave amplitude, which does not permit the stiff matter 
period to extend all the way to nucleosynthesis \cite{SSS}. In practice 
therefore the 
increase permitted will not be as large as this calculation indicates. However, 
a rather detailed calculation would be required to determine the balance of 
reducing the inflationary energy 
scale and shortening the stiff matter era which maximizes $N$ without violating 
the gravitational wave constraint.

\subsection{Early matter domination}

One possible modification to the simplest cosmology is for the long 
radiation-dominated epoch after reheating to be punctuated by epochs of 
matter domination, for example when long-lived massive particles go out of 
equilibrium and come to dominate the Universe before decaying. Moduli fields 
provide an example \cite{moduli}, though they are too long-lived in many 
scenarios to be compatible with requirements.

Inserting a period of matter domination into Eq.~(\ref{e:efolds}) is simple, and 
it reduces $N_{{\rm hor}}$ by $\Delta N = [\ln \rho_{{\rm f}}/\rho_{{\rm 
i}}]/12$ 
where $\rho_{{\rm i}}$ and $\rho_{{\rm f}}$ are the densities at the beginning 
and end of the matter-dominated era, just as in the derivation of 
Eq.~(\ref{e:standard}). A very prolonged period of matter domination is required 
to give a significant reduction.

\subsection{Thermal inflation}

Thermal inflation was introduced in Ref.~\cite{thermal} as a means of solving 
relic abundance problems left over from the original phase of inflation. It is 
envisaged as one or more short periods of inflation, which are not so prolonged 
as to generate observable perturbations. The consequence pertinent to the 
present discussion is that thermal inflation corresponds to an extra stretching 
of the primordial perturbations, thus reducing $N_{{\rm hor}}$.

Under the reasonable assumption that the energy density does not change 
significantly during thermal inflation, the effect is simply to reduce $N_{{\rm 
hor}}$ by the number of $e$-foldings $N_{{\rm thermal}}$ of thermal inflation. 
If thermal inflation is to achieve its purpose, this number is expected to be 
about 10, though there is also the possibility of multiple periods of thermal 
inflation.

\subsection{Braneworld cosmology}

Another possible modification to the standard cosmology is if the Friedmann 
equation is modified at high energies, the archetypal example being the 
braneworld cosmology. For example, in the Randall--Sundrum Type II model 
\cite{RSII}, at high energies we expect
\begin{equation}
H^2 = \frac{8\pi}{3 m_{{\rm Pl}}^2} \left( \rho + \frac{\rho^2}{2\lambda}
	\right) \,,
\end{equation}
where $\lambda$ is the brane tension. A full discussion of the consequences of 
this is beyond the scope of this paper, but we note an interesting case where 
$\lambda$ is much smaller than the energy at the end of inflation, so that the 
initial phase of the reheating, and possibly of the radiation-dominated era, 
take place during the high-energy regime $\rho \gg \lambda$.

Within the high-energy regime, the expansion laws corresponding to matter and 
radiation domination are slower than in the standard cosmology, being $a \propto 
t^{1/3}$ and $a \propto t^{1/4}$ respectively, though the behaviour of the 
densities as a function of the scale factor is unchanged. Slower expansion rates 
mean a 
greater change in $aH$ relative to the change in $a$, which can increase 
$N_{{\rm hor}}$. However a full 
calculation would have to include that inflation was taking place during the 
high-energy regime, which tends to force down the normalization of the potential 
giving rise to a particular amplitude of perturbations \cite{MWBH}, and is 
beyond the scope of this paper.

\subsection{An absolute minimum for $N_{{\rm hor}}$}

Given the uncertainties in the cosmological model, is it possible to say 
anything robust concerning the minimum possible value of $N_{{\rm hor}}$? The 
only guidance is that the success of primordial nucleosynthesis suggests that we 
should not seek to modify the standard cosmology after that epoch. As 
\begin{equation}
\frac{a_{{\rm nuc}} H_{{\rm nuc}}}{a_0 H_0} \simeq 10^8 \,,
\end{equation}
we conclude that $N_{{\rm hor}}$ has a minimum of about 18 $e$-foldings from the 
end of inflation. However, this extreme limit can only be realized in the 
unlikely case that either all the inflation really happened at such a low scale, 
or where repeated bouts of thermal inflation served to hold the perturbations on 
superhorizon scales long after they were formed. 

\section{Conclusions}

We have carried out an extensive analysis seeking to clarify the appropriate 
choices for the number of $e$-foldings from the end of inflation corresponding 
to observed perturbations. Assuming the simplest cosmology, we find a plausible 
maximum value of around 60, in good agreement with a recent paper of Dodelson 
and Hui \cite{DH}, but noted that even fairly standard scenarios can violate it, 
an example being the $\lambda \phi^4$ case which gives a higher value of 64 
$e$-foldings. That model is also an 
exceptional one where a more accurate calculation is possible despite 
uncertainties about reheating.

In general, however, the number is sensitive both to a possible reduction in 
energy scale during the late stages of inflation, and to the complete 
cosmological 
evolution, and we have highlighted the effects of some plausible non-standard 
scenarios. In some cases, these may permit a higher maximum number of 
$e$-foldings than the standard cosmology.

Obviously the total number of $e$-foldings of inflation must be greater than 
$N_{{\rm hor}}$, which concerns only observable scales. In almost all models of 
inflation it is expected to be very much greater, though these $e$-foldings are 
not accessible to observations.

In summary, for a typical inflation model it remains a sensible working 
hypothesis that the number of 
$e$-foldings lies between 50 and 60, where this number refers to the amount of 
expansion from when our present Hubble radius equalled the Hubble radius during 
inflation up until the end of inflation. However, if a particular model is under 
investigation, it may pay to attempt a more accurate calculation, at least to 
highlight the effect of assumptions concerning the cosmological evolution. This 
is particularly true if the model is expected to have a slow rate of inflation 
at its late stages, or to have an unusually low energy scale (corresponding to 
very small $\epsilon$), or to have a particularly prolonged reheating period.


\begin{acknowledgments}
S.M.L.~was support by the European Union CMBNET network, and A.R.L.~in part by 
the Leverhulme Trust. We thank Micha\"el Malquarti for useful discussions.
\end{acknowledgments}


\end{document}